\documentclass[final,times,fleqn]{elsarticle}

\usepackage{natbib}
\usepackage{graphicx}
\usepackage{amsmath}
\usepackage{amssymb}
\usepackage{amsfonts}
\usepackage[%
]{hyperref}
\usepackage{mathrsfs}
\usepackage{dcolumn}

\newcommand{\re}{\mathop{\mathrm{Re}}}
\newcommand{\im}{\mathop{\mathrm{Im}}}
\journal{Physics Letters B}

\begin{document}

\begin{frontmatter}

\title{Symmetries of higher-spin fields\\and the electromagnetic $N \to N^*(1680)$ form factors}

\author[SFU,INR]{G.~Vereshkov}
	\ead{gveresh@gmail.com}
	
\author[SFU]{N.~Volchanskiy\corref{corauth}}
	\ead{nikolay.volchanskiy@gmail.com}
	
\address[SFU]{%
Research Institute of Physics,
Southern Federal University,
Prospekt Stachki, 194, 344090 Rostov-na-Donu, Russia
}

\address[INR]{%
Institute for Nuclear Research of the Russian Academy of Sciences,
Prospekt 60-letiya Oktyabrya, 7a, 117312 Moscow, Russia
}

\cortext[corauth]{Corresponding author}

\begin{abstract}
We study the $Q^2$-evolution of the form factors (FFs) for the nucleon-to-$N^*(1680)$ transition in the framework of an effective field theory. To this end, the intrinsic symmetries of the spin-$\frac52$ Rarita-Schwinger (RS) fields are analyzed, and a Lagrangian of the electromagnetic $NN^*(1680)$ interactions is constructed. The Lagrangian preserves all the intrinsic symmetries of the spin-$\frac52$ field---point and gauge invariance---and does not involve lower-spin components of the reducible RS field. Besides, the symmetries postulate the definitions of the Lagrangian FFs. These FFs are modeled as dispersionlike expansions in a vector-meson--dominance model. A good agreement with the experimental data is achieved.
\end{abstract}


\end{frontmatter}


\section{Introduction}

The resonance $N^*(1680)$ enjoys a special place in the family of nonstrange baryons. Carrying as it does a spin-parity $J^P = \left(\frac52\right)^+$, the $N^*(1680)$ is the highest-spin nucleon resonance, all three nucleon-to-resonance form factors (FFs) of which can be reliably extracted from the current experimental data \cite{2007drechsel, az-05b}. However, to our knowledge, in the literature there is no Lagrangian effective-field model of the $N \to N^*(1680)$ electromagnetic interactions such that the $N^*(1680)$ field involves correct number of degrees of freedom (DsOF) both on-shell and off-shell.

In this letter we construct a Lagrangian of the electromagnetic $NN^*(1680)$ interactions that preserves all the symmetries of the free spin-$\frac52$ Rarita-Schwinger (RS) field. Such interactions involve only physical DsOF of the reducible field. Using this Lagrangian we study the $Q^2$-evolution of the helicity amplitudes for the transition $N \to N^*(1680)$. The remainder of the paper is organized as follows.

In the Section \ref{sec:symmetries} we recover the Lagrangian of the free spin-$\frac52$ RS field \cite{2009shklyar}. Here we show that it possesses symmetry properties similar to those of the spin-$\frac32$ RS field---invariance under point and gauge transformations of the field. These intrinsic symmetries of the field are linked to the constraints eliminating redundant lower-spin components of the reducible RS field. To obtain consistent interacting spin-$\frac52$ theory, we should require the invariance of the interaction Lagrangian under the free-field symmetries.

The $\gamma^*NN^*(1680)$-Lagrangian preserving the required symmetries is written down in the Section \ref{sec:lagr}. Three terms of the minimally local Lagrangian turn out to be defined uniquely and involve different numbers of the field derivatives, which provides us with the classification of the FFs in terms of the differential order of the Lagrangian. Besides, we find that the Lagrangian shares much in common with the $\gamma^*N\Delta(1232)$-Lagrangian \cite{2009vereshkov} of the same symmetry in its form and properties.

In the Section \ref{sec:HAs} the helicity amplitudes of the electromagnetic $NN^*(1680)$-transition are calculated using the constructed Lagrangian. The relations between the helicity amplitudes and the Lagrangian FFs prove to be the same as in the case of the spin-$\frac32$ resonance. It suggests that the symmetries considered provide us with a unified description of the interactions of the higher-spin resonances. Also in this Section the high-$Q^2$ behavior of the amplitudes and the FFs is considered.

In the Section \ref{sec:fits} the available experimental data is fitted using the formulas of the model developed. Finally, the Section \ref{sec:conclusion} is a summary of the results of the paper. Here we also stress many similarities between the point and gauge invariant theories of the transitions $N \to \text{spin-$\frac32$ and spin-$\frac52$}$ resonances.


\section{\label{sec:symmetries}Internal symmetries of the spin-$\frac52$ fields}

The Lagrangian of the free symmetric Rarita-Schwinger (RS) field $\Psi_{\mu\nu}$ should lead to the Dirac equation for each tensor component of the field and the constraints eliminating redundant DsOF of the reducible field $\Psi_{\mu\nu}$ \cite{ra-sh},
\begin{align}
	\left( i \gamma_\lambda \partial^\lambda - M \right) \Psi_{\mu\nu} = 0,
	\\ \label{constraints}
	\partial^\lambda \Psi_{\lambda\mu} = 0 = \gamma^\lambda \Psi_{\lambda\mu}.
\end{align}
To obtain the constraints for the RS field with spin $J \geqslant \frac52$, it is always necessary to introduce auxiliary fields \cite{1979berends}. The most general Lagrangian for the spin-$\frac52$ can be written as
\begin{align}\label{L52}
	\notag\mathscr{L}_\text{ff}
	 = {}&\bar\Psi^{\mu\nu} \left[ i \Gamma^{(1)}_{\{\mu\nu\}\{\lambda\sigma\}\rho}(A) \partial^\rho - M \Gamma^{(1)}_{\{\mu\nu\}\{\lambda\sigma\}}(A) \right] \Psi^{\lambda\sigma}
	 \\\notag
	 &{}+\bar\Psi^{\mu\nu} \left[ i \Gamma^{(2)}_{(\mu\nu)(\lambda\sigma)\rho}(A,B) \partial^\rho - M \Gamma^{(2)}_{(\mu\nu)(\lambda\sigma)}(A,B) \right] \Psi^{\lambda\sigma}
	 \\
	 &{}+\kappa M \bar\Psi_\lambda^\lambda \Theta - \frac{3\lvert 3A-2 \rvert^2 \kappa^2}{80 \lvert B \rvert^2} \bar\Theta \left( i\gamma_\lambda \partial^\lambda +3M\right) \Theta
	 + \text{ H.c.}
\end{align}
Here $\Theta(x)$ is an auxiliary spin-$\frac12$ field, $\kappa$ is an arbitrary real parameter, $A\neq\frac23$, $B\neq 0$ are complex parameters, and the round brackets $(\cdot\cdot)$ enclose symmetric pairs of indices, $A_{(\mu\nu)} = \frac12 (A_{\mu\nu} + A_{\nu\mu})$, while the curly ones $\{\cdot\cdot\}$ enclose symmetric and traceless pairs, $A_{\{\mu\nu\}} = A_{(\mu\nu)} - \frac14 A_\mu{}^\mu$. The tensor spinors in the kinetic and mass terms of the Lagrangian \eqref{L52} are given by
\begin{align*}\label{}
	\Gamma^{(1)}_{\mu\nu\lambda\sigma\rho}(A) =
	\Biggl[ g_{\mu\lambda} \gamma_\rho - A g_{\mu\rho} \gamma_\lambda - A^* g_{\lambda\rho} \gamma_\mu
	+ \frac12 \left( \frac54 \lvert A \rvert^2 - \re A + 1 \right)
	\gamma_\mu \gamma_\rho \gamma_\lambda \Biggr] g_{\nu\sigma},
\end{align*}
\begin{align*}\label{}
	\Gamma^{(1)}_{\mu\nu\lambda\sigma}(A) = 
	\Biggl[ g_{\mu\lambda} -\frac1{4} \left( \frac{15}{4} \lvert A \rvert^2 - 5 \re A + 3 \right) \gamma_\mu \gamma_\lambda \Biggr] g_{\nu\sigma},
\end{align*}
\begin{align*}\label{}
	\Gamma^{(2)}_{\mu\nu\lambda\sigma\rho}(A,B) =
	-\frac{16 \lvert B \rvert^2 + \lvert 3A-2 \rvert^2 \re B}{\lvert 3A-2 \rvert^2} g_{\mu\nu} g_{\lambda\sigma} \gamma_\rho
	+ 2B g_{\lambda\sigma} g_{\nu\rho} \gamma_\mu + 2B^* g_{\mu\nu} g_{\sigma\rho} \gamma_\lambda,
\end{align*}
\begin{align*}\label{}
	\Gamma^{(2)}_{\mu\nu\lambda\sigma}(A,B) =
	-\frac{48 \lvert B \rvert^2}{\lvert 3A-2 \rvert^2} g_{\mu\nu} g_{\lambda\sigma}.
\end{align*}
For the real values of the free-field parameters, $\im A = \im B = 0$, the Lagrangian \eqref{L52} was derived in Ref. \cite{2009shklyar}.

Evaluating field equations for the Lagrangian \eqref{L52} and operating on them by $\gamma^\mu$ and $\partial^\mu$ give all the necessary constraints and eliminate the auxiliary field, $\Theta = 0$. We thereby are led to the conclusion that the free RS field equations are derived from a two-parameter equivalent class of actions $S_\text{ff} \left(A,\,B\right)$.

The spin-$\frac52$ RS field possesses the similar intrinsic symmetries as in the case of spin $\frac32$. In the massless limit the Lagrangian is invariant under gauge transformation of the field,
\begin{align}\label{gt}
	\Psi_{\mu\nu} \to \Psi_{\mu\nu}
	 +\frac12 \left( \partial_\mu \alpha_\nu + \partial_\nu \alpha_\mu \right),
\end{align}
where $\alpha_\mu$ is a vector spinor field subjected to conditions $\partial^\lambda \alpha_\lambda = 0 = \gamma^\lambda \alpha_\lambda$ and $\gamma_\lambda \partial^\lambda \alpha_\mu = 0$ (The latter condition is unnecessary for $A=2$.). Besides, the equivalent class of the Lagrangians \eqref{L52} is invariant under two-parameter point transformation,
\begin{align}\label{pt}
	\notag&\Psi_{\mu\nu} \to \Lambda_{\mu\nu}{}^{\lambda\sigma} (A_2,B_2 \vert A_1,B_1) \Psi_{\lambda\sigma},
	\\
	&\Theta \to \frac{B_1^*}{B_2^*} \frac{3A_2-2}{3A_1-2} \Theta, 
\end{align}
where
\begin{align*}\label{}
	\Lambda_{\mu\nu}{}^{\lambda\sigma} (A_2,B_2 \vert A_1,B_1) = 
	\delta_{\{\mu}^{\{\lambda} \delta_{\nu\}}^{\sigma\}}
	+\frac{A_2-A_1}{3 A_1 - 2} \gamma_{\{\mu} \gamma^{\{\lambda} \delta_{\nu\}}{}^{\sigma\}} +\frac14 \frac{B_2}{B_1} \frac{3A^*_1-2}{3A^*_2-2} g_{\mu\nu} g_{\lambda\sigma}.
\end{align*}
The point transformation \eqref{pt} shifts the parameters of the free-field Lagrangian \eqref{L52} from the values $A_1$, $B_1$ to $A_2$, $B_2$. As in the case of the spin-$\frac32$ RS field \cite{2004pilling, 2005pilling} the point transformations \eqref{pt} form a non-unitary symmetry group, with the multiplicative law being given by
\begin{align*}\label{}
	\Lambda_{\mu\nu}{}^{\rho\omega} (A_2,B_2 \vert A_0,B_0) \Lambda_{\rho\omega}{}^{\lambda\sigma} (A_0,B_0 \vert A_1,B_1) = \Lambda_{\mu\nu}{}^{\lambda\sigma} (A_2,B_2 \vert A_1,B_1)
\end{align*}


\section{\label{sec:lagr}Lagrangian of the electromagnetic $NN^*(1680)$-interactions}

The consistent interaction Lagrangian of the spin-$\frac52$ field should involve only physical DsOF of the reducible RS field. Put in other words, the Green's function sandwiched between vertex operators should reduce to the spin-$\frac52$ projection operator,
\begin{align*}\label{}
	J^{\mu\nu}_{(1)} G_{\mu\nu\lambda\sigma}(p) J^{\lambda\sigma}_{(2)} =
	J^{\mu\nu}_{(1)} \frac1{\hat p - M + i 0} P^{(5/2)}_{\mu\nu\lambda\sigma}(p) J^{\lambda\sigma}_{(2)},
\end{align*}
where $J^{\mu\nu}_{(1,2)}$ are some tensor-spinor currents. The projection operator $P^{(5/2)}_{\mu\nu\lambda\sigma}(p)$ \cite{1979berends} can be written as
	\begin{align}\label{}
	P^{(5/2)}_{\mu\nu\lambda\sigma}(p) = \frac1{(p^2)^2} \Gamma_{\mu\alpha\nu\beta\lambda\gamma\sigma\delta} p^\alpha p^\beta p^\gamma p^\delta,
\end{align}
where the tensor spinor $\Gamma_{\mu\alpha\nu\beta\lambda\gamma\sigma\delta}$ is introduced below in Eq.~\eqref{Gamma8}. The projector is $\gamma$- and $p$-transversal,
\begin{align*}\label{}
	&\gamma^\mu P^{(5/2)}_{\mu\nu\lambda\sigma}(p) = 0 = p^\mu P^{(5/2)}_{\mu\nu\lambda\sigma}(p),
	\quad
	P^{(5/2)}_{\mu\nu\lambda\sigma}(p) = P^{(5/2)}_{\nu\mu\lambda\sigma}(p),
	\\
	&P^{(5/2)}_{\mu\nu\lambda\sigma}(p) \gamma^\lambda = 0 = P^{(5/2)}_{\mu\nu\lambda\sigma}(p) p^\lambda,
	\quad
	P^{(5/2)}_{\mu\nu\lambda\sigma}(p) = P^{(5/2)}_{\nu\mu\sigma\lambda}(p).
\end{align*}

It can be shown that there are two classes of such interactions. \textit{The first class} is the interactions invariant under the gauge transformations \eqref{gt} and $B$-part of the point transformations \eqref{pt}. It is a direct generalization of the spin-$\frac32$ interactions studied in Refs.~\cite{1998pascalutsa, pa-ti, pa-07}. Such interactions do not involve lower-spin components of the RS field only for the exceptional value of the free-field parameter $A=2$ and could modify the constraints \eqref{constraints} making them nonlinear. Therefore, the first-class interactions break the point invariance of the equivalent class of the Lagrangians \eqref{L52}. The gauge-invariant interactions are described by 4-transversal traceless tensor-spinor currents,
\begin{align*}\label{}
	\partial^\mu J_{\mu\nu} = 0, \quad J_\mu^\mu = 0,
	\qquad
	J_{\mu\nu} = \frac{\delta}{\delta \bar\Psi^{\mu\nu}} \int d^4 x \mathscr{L}_\text{int}(x).
\end{align*}
\textit{The second class} is the interactions that are invariant under both the gauge and point transformations of Eqs. \eqref{gt} and \eqref{pt}. Such interactions are described by 4- and $\gamma$-transversal currents, $\partial^\mu J_{\mu\nu} = 0 = \gamma^\mu J_{\mu\nu}$. The point and gauge invariant interaction Lagrangians is a subclass of gauge invariant ones. They are different, however, in that the point and gauge invariant interactions leave the free-field constraints \eqref{constraints} intact. The point and gauge invariant Lagrangians were considered in Refs.~\cite{2009vereshkov} and \cite{2008shklyar, 2009shklyar} for the cases of electromagnetic $N\Delta(1232)$-interactions and trilinear interactions of pions and nucleons with spin-$\frac32\left(\frac52\right)$ resonances, respectively.

It is easy to see that the first class of interactions includes infinite number of possible interaction Lagrangians. Fortunately, the point and gauge invariant interactions of the second class imply more strict constraints on the symmetry of the Lagrangian. Preserving the simple structure of the free-field linear constraints, the point and gauge invariant interactions result in the Lagrangian that is defined unambiguously. Therefore, in what follows we will consider interactions that are both point and gauge invariant.

The most general gauge and point invariant local Lagrangian of $\gamma^* N N^*(1680)$-inter\-actions is as follows,
\begin{align}\label{L123}
	&\notag\mathscr{L}_{(1)}
	 =
	 \sum_V \frac{ig_1^V}{2M_N^3 M_R^2} \bar\Psi^{AB,\alpha} \Gamma_{ABC\alpha\beta} N^{,\beta} V^{C}
	 + \text{ H.c.},
	 \\
	&\notag\mathscr{L}_{(2)}
	 =
	 \sum_V \frac{g_2^V}{2M_N^3 M_R^3} \bar\Psi^{AB,\alpha\zeta} \Gamma_{ABC\alpha\beta} \gamma_{\zeta} N^{,\beta} V^{C} + \text{ H.c.},
	 \\
	&\notag\mathscr{L}_{(3)}
	 =
	 \sum_V \frac{ig_3^V}{2M_N^3 M_R^4} \bar\Psi^{AB,\alpha\zeta}
	 \Bigl( \Gamma_{AB\rho\zeta\alpha\beta} g_{\omega\eta} 
	 -\Gamma_{AB\omega\zeta\alpha\beta} g_{\rho\eta}
	 \\ &\phantom{\mathscr{L}_{(3)}={}}
	 +\Gamma_{AB\rho\eta\alpha\beta} g_{\omega\zeta} 
	 -\Gamma_{AB\omega\eta\alpha\beta} g_{\rho\zeta}
	 -\Gamma_{AB\rho\omega\alpha\beta} g_{\zeta\eta} \Bigr)
	 N^{,\beta\eta} V^{\rho\omega} + \text{ H.c.}
\end{align}
Here capital Latin letters denote bi-indices, $V_{\mu\nu}$ are strength tensors of photon and vector-meson fields, and $\Psi_{\mu\nu\lambda\sigma}$ is the gauge invariant operator which shares the properties of the Riemann–--Christoffel tensor,
\begin{align*}\label{}
	&\Psi_{\mu\nu\lambda\sigma} = \frac12 \left( \partial_\nu \partial_\lambda \Psi_{\mu\sigma} + \partial_\mu \partial_\sigma \Psi_{\nu\lambda} - \partial_\nu \partial_\sigma \Psi_{\mu\lambda} - \partial_\mu \partial_\lambda \Psi_{\nu\sigma} \right),
	\\
	&\Psi_{\mu\nu\lambda\sigma} = -\Psi_{\nu\mu\lambda\sigma} = -\Psi_{\mu\nu\sigma\lambda},
	\qquad
	\Psi_{\mu\nu\lambda\sigma} = \Psi_{\lambda\sigma\mu\nu}.
\end{align*}
The 8th-rank coupling tensor spinor $\Gamma_{ABCD}$ is antisymmetric under permutations in the four pairs of its indices, $\Gamma_{ABCD}
= -\Gamma_{\bar{A}BCD} = -\Gamma_{A\bar{B}CD} = -\Gamma_{AB\bar{C}D} = -\Gamma_{ABC\bar{D}}$ (bars denote permutation of the indices, $A=\mu\nu$ and $\bar A = \nu\mu$), and symmetric under permutations of the first and last two pairs, $\Gamma_{ABCD} = \Gamma_{BACD} = \Gamma_{ABDC}$. Due to the point invariance of the interaction Lagrangian, the coupling tensor spinor is $\gamma$-transversal, $\gamma^\mu \Gamma_{\mu\nu BCD} = 0 = \Gamma_{ABC\alpha\beta} \gamma^\beta$. The coupling tensor spinor can be explicitly written as
\begin{align}\label{Gamma8}
	\notag\Gamma_{ABCD}
	={}&
	-\frac9{10} \left( \Gamma_{AD} \Gamma_{BC}
	+\Gamma_{BC} \Gamma_{AD}
	+\Gamma_{AC} \Gamma_{BD}
	+\Gamma_{BD} \Gamma_{AC} \right)
	\\\notag&
	+\frac3{10} \left( \Gamma_{AB} \Gamma_{DC}
	+\Gamma_{BA} \Gamma_{DC}
	+\Gamma_{AB} \Gamma_{CD}
	+\Gamma_{BA} \Gamma_{CD} \right)
	\\&
	-\frac3{10} \left( \Gamma_{AC} \Gamma_{DB}
	+\Gamma_{BC} \Gamma_{DA}
	+\Gamma_{AD} \Gamma_{CB}
	+\Gamma_{BD} \Gamma_{CA} \right),
\end{align}
where
\begin{align}\label{kernel}
	\Gamma_{\mu\nu\lambda\sigma}
	=
	\frac13 \biggl[ e_{\mu\nu\lambda\sigma} +
	i \gamma_5 \left(g_{\mu\lambda} g_{\nu\sigma} - g_{\nu\lambda} g_{\mu\sigma} \right)
	-\frac12 \left( g_{\mu\lambda} \tilde\sigma_{\nu\sigma}
	                - g_{\nu\lambda} \tilde\sigma_{\mu\sigma}
	                - g_{\mu\sigma} \tilde\sigma_{\nu\lambda}
	                + g_{\nu\sigma} \tilde\sigma_{\mu\lambda} \right)
	\biggr].
\end{align}
Here $\tilde\sigma_{\mu\nu} = \frac12 e_{\mu\nu\eta\xi} \sigma^{\eta\xi}$ is dual to $\sigma_{\mu\nu} = \frac12 (\gamma_\mu \gamma_\nu - \gamma_\nu \gamma_\mu)$.

It should be noted that the Lagrangian \eqref{L123} is defined uniquely in each given differential order. This follows from the fact that any $\gamma$-transversal coupling tensor spinors $\Gamma_{\mu\nu\lambda\sigma\dots}$ permitted by the symmetry of the theory contain the universal 8th-rank tensor spinor $\Gamma_{ABCD}$ as a carrier of its symmetry properties,
\begin{align}\label{}
	\Gamma_{AB\dots} = \Gamma_{AB}{}^{CD} \left( R_{CD\dots} + \gamma^\eta S_{CD\eta\dots} + \sigma^{\eta\zeta} T_{CD\eta\zeta\dots} \right),
\end{align}
where $R$, $S$, $T$ are some coefficient tensors.

Another attractive property offered by the point and gauge invariant Lagrangian \eqref{L123} is that it has the same structure as the point and gauge invariant Lagrangian in the case of spin-$\frac32$ RS field \cite{2009vereshkov},
\begin{equation}\label{L V}
\begin{aligned}
	\mathscr{L}_{(1)} ={}&
		\sum_V \frac{i g_{1}^V}{2M_N^2} \bar \Psi^{\mu\nu}
		\Gamma_{\mu\nu\lambda\sigma} V^{\lambda\sigma} N + \text{H.c.},
	\\
	\mathscr{L}_{(2)}
	={}& -\sum_V \frac{g_{2}^V}{2 M_N^2 M_R^{\hphantom{2}}} \bar \Psi^{\mu\nu,\omega} \Gamma_{\mu\nu\lambda\sigma} \gamma_\omega V^{\lambda\sigma} N + \text{H.c.},
	\\
	\mathscr{L}_{(3)} = {}&
		\sum_V \frac{i g_{3}^V}{2M_N^2 M_R^2} \bar \Psi^{\mu\nu,\rho}
		\bigl( \Gamma_{\mu\nu\lambda\rho} g_{\sigma\omega}
		      -\Gamma_{\mu\nu\sigma\rho} g_{\lambda\omega}
		      \\&
		      +\Gamma_{\mu\nu\lambda\omega} g_{\sigma\rho}
		      -\Gamma_{\mu\nu\sigma\omega} g_{\lambda\rho}
		      -\Gamma_{\mu\nu\lambda\sigma} g_{\rho\omega}
		\bigr) V^{\lambda\sigma} N^{,\omega}
	+ \text{H.c.},
\end{aligned}
\end{equation}
where $\Psi_{\mu\nu} = \partial_\mu \Psi_\nu - \partial_\nu \Psi_\mu$ is the field strength tensor. Comparing the Lagrangians \eqref{L123} and \eqref{L V} it is readily seen that they share all mathematical properties---(i) in both cases there exists the universal $\gamma$-transversal coupling tensor spinor; (ii) relations between higher--derivative couplings and the universal tensor spinor has the same structure in both Lagrangians; (iii) all three local couplings are defined uniquely and classified in terms of the differential order of the corresponding Lagrangian. Therefore, the point and gauge invariance of the Lagrangian result in the unified structure of the Lagrangians, at least in the cases of spin-$\frac32$ and $\frac52$. It is also important that the 8th-rank coupling tensor spinor \eqref{Gamma8} for spin-$\frac52$ is related to the 4th-rank coupling \eqref{kernel} for spin-$\frac32$. It can be expected that this is the regular pattern that will remain for all higher-spin resonances $J \geqslant \frac72$.


\section{\label{sec:HAs}Helicity amplitudes and form factors}

The helicity amplitudes for the electroproduction of the $N^*(1680)$ resonance on the mass shell calculated using the point and gauge invariant Lagrangian \eqref{L123} are
\begin{align}\label{A32}
	A_{3/2}
	{}&= \sqrt{N} \Bigl[ \left( Q^2 - \mu M_N \right) F_1
	+ \mu M_R F_2 - \left( Q^2+ \mu M_R \right) F_3 \Bigr],
	\\ \label{A12}
	A_{1/2}
	{}&= -\sqrt{\frac{N}2} \Bigl[ \mu M_R F_1
	+ \left( Q^2 - \mu M_N \right) F_2 + \mu M_N F_3 \Bigr],
	\\ \label{S12}
	S_{1/2}
	{}&= \sqrt{\frac{N}4} Q_+ Q_- \Biggl[ F_1 - F_2
	+ \frac{Q^2+M_R^2+M_N^2}{2M_R^2} F_3 \Biggr],
\end{align}
where $F_f = F_f(Q^2)$, $f = 1,\, 2,\, 3$ are the point and gauge invariant FFs, $\mu=M_R-M_N$, $N=\frac{2\pi \alpha Q_-^2 Q_+^4}{5 M_N^9 \left( M_R^2 - M_N^2 \right)}$, $Q_{\pm} = \left[ Q^2 + (M_R \pm M_N)^2 \right]^{-1/2}$.

In Eqs.~\eqref{A32}--\eqref{S12} there are two types of the $Q^2$-dependent functions (except FFs themselves). The functions of the first type are square-root common factors that are universal kinematic factors present in any Lagrangian model \cite{Devenish1976}. The second type of functions is polynomials accompanying FFs. The particular form of these polynomials depends upon the choice of the Lagrangian. It is interesting to note that the polynomials in Eqs.~\eqref{A32}--\eqref{S12} are the same as in the case of the spin-$\frac32 $ resonance $\Delta(1232)$ (see Eqs.~(6)--(8) in \cite{2009vereshkov}) up to the change $M_R \to -M_R$, since the resonance $\Delta(1232)$ is of abnormal parity, $J^P = \frac32^+$. It means that the point and gauge invariance produces the universal structure of the electroproduction helicity amplitudes, with the spin of the excited resonance being irrelevant. It should be stressed that such simple result for observables is obtained, despite the fact that the spin-$\frac52$ coupling \eqref{Gamma8} is much more lengthy than the spin-$\frac32$ one \eqref{kernel}. The independence of the amplitudes in the spin of the resonance results from the existence of the universal $\gamma$-transversal structure \eqref{kernel}---any coupling in a point and gauge invariant Lagrangian is expressed through the tensor spinor \eqref{kernel}.

At asymptotically high $Q^2$ pQCD predicts the scaling behavior of the helicity amplitudes \eqref{A32}--\eqref{S12} to be \cite{st-93}
\begin{equation}\label{HA scaling}
	A_{3/2} \sim \frac{1}{Q^5 \ell^{n_1}}, \quad
	A_{1/2} \sim \frac{1}{Q^3 \ell^{n_2}}, \quad
	S_{1/2} \sim \frac{1}{Q^3 \ell^{n_3}},
\end{equation}
where $\ell = \ln{(Q^2/\Lambda^2)}$ and $n_2-n_3 \approx 2$.
From Eqs.~\eqref{A32}--\eqref{S12} and \eqref{HA scaling} it follows that the high-$Q^2$ behavior of the FFs is
\begin{align}\label{FF scaling}
	F_1 \sim \frac{1}{Q^{10} \ell^{n_1}}, \quad
	F_2 \sim \frac{1}{Q^8 \ell^{n_2}}, \quad
	F_3 \sim \frac{1}{Q^{10} \ell^{n_3}}, \quad
\end{align}
and $n_3>n_1$. This implies that the FFs $F_1(Q^2) \sim Q^{-5} A_{3/2}$, $F_2(Q^2) \sim Q^{-5} A_{1/2}$, $F_3(Q^2) \sim Q^{-7} S_{1/2}$ acquire (in the asymptotic domain) the statuses of, respectively, the FF of the processes involving flips of two quark helicities, the non-helicity-flip FF, and the helicity-flip FF.

It should be noted that such high-$Q^2$ properties of the FFs are naturally consistent with the classification of the FFs in terms of the differential order of the interaction Lagrangian. Indeed, the first and the third terms of the Lagrangian \eqref{L123} involve baryon fields of opposite chiralities and, thus, describe electroproduction with the flip of baryon helicity. The second term, contrarily, links the nucleon and resonance fields of the same chirality and, consequently, amounts to helicity-conserving interactions.

Within the vector-meson--dominance (VMD) model \cite{vereshkov:073007}, the FFs $F_f(Q^2)$ are given by dispersionlike expansions
\begin{align}\label{FFs}
	F_f^{p,n}(Q^2) = \sum_{k=1}^{K} \frac{m_k^2\varkappa^{p,n}_{kf}(Q^2)}{m_k^2+Q^2}
	 = \sum_{n=0}^{\infty} \frac{(-1)^n}{Q^{2n}}
	   \sum_{k=1}^{K}{m_k^{2n}\varkappa_{kf}^{p,n}(Q^2)},
\end{align}
with the poles being at the masses $m_k$ of the vector mesons listed in Tab.~\ref{tab:mesons}. Note that, since the $\omega$ and $\rho$-mesons form singlet-triplet families with near mass degeneracy in the families, we neglect singlet-triplet mass splitting and introduce averaged masses $m_k^2 = (m_{(\omega)k}^2 + m_{(\rho)k}^2)/2$. Besides, in Eq.~\eqref{FFs} it is supposed that $\omega$ and $\rho$-mesons propagate in the nucleon medium identically, i.e. $\varkappa^{\omega}_{kf}(Q^2)/\varkappa^{\omega}_{kf}(0) = \varkappa^{\rho}_{kf}(Q^2)/\varkappa^{\rho}_{kf}(0)$. This is justified by the lack of the data on the neutron FFs for the transition $n\gamma^*\to N^*_{0}(1680)$.

To assure correct high-$Q^2$ behavior of the dispersionlike expansions of the FFs \eqref{FFs}, we assume the following: (i) The $Q^2$-dependence of the expansion coefficients is independent of the meson-family index $k$, $\varkappa_{kf}(Q^2) = \varkappa_{kf}(0)/L_f(Q^2)$.
(ii) The interpolation functions $L_f(Q^2)$ are given by $L_f = \left( 1 + b_f \bar\ell + a_f \bar\ell^2 \right)^{n_f/2}$, $\bar\ell = \ln\left( 1+{Q^2}/{\Lambda^2} \right)$, which effectively takes into account the renormalization of the strong coupling constant and the $Q^2$-evolution of the parton distribution functions \cite{1973Brodsky-Farrar,1975Brodsky-Farrar}.
(iii) The parameters of the meson spectrum satisfy the superconvergence relations $\sum_k m_k^{2n} \varkappa_{kf}^{\vphantom2}(0) =0$ for $f=1,\,2,\,3$, $n=1,\,2,\,3$ and $f=1,\,3$, $n=4$.

\begin{table}
\caption{\label{tab:mesons}PDG vector-meson masses (GeV) \cite{r2008PDG}. The isosinglet mesons $\omega(1960)$ and $\omega(2205)$ are listed in the section ``Further states'' \cite{r2008PDG}. The last column gives an averaged mass $m_k = \left[ (m_{(\omega)k}^2 + m_{(\rho)k}^2)/2 \right]^{1/2}$.}
\small
\begin{tabular}{cccccc}
$k$&$$&$m_{(\rho)k}$ &&$m_{(\omega)k}$&$m_k$\\
\hline
$1$ & $\rho(770)$ & 0.77549 & $\omega(782)$ & 0.78265 & $0.77908$ \\
$2$ & $\rho(1450)$ & 1.465& $\omega(1420)$ & 1.425& $1.445$ \\
$3$ & $\rho(1700)$ & 1.720& $\omega(1650)$ & 1.670& $1.695$ \\
$4$ & $\rho(1900)$ & 1.885& $\omega(1960)$ & 1.960& $1.923$ \\
$5$ & $\rho(2150)$ & 2.149& $\omega(2205)$ & 2.205& $2.177$ \\
\end{tabular}
\end{table}


\section{\label{sec:fits}Data analysis}

\begin{table}
\caption{\label{tab:param}Fit parameters.}
\small
\begin{tabular}{cD{.}{.}{-1}ccD{.}{.}{-1}}
$\varkappa^p_{52}(0)$ & 0.1383 &\qquad\quad& $a_1$ & 0.6242 \\
$\sum_{k=1}^5 \varkappa^p_{k1}(0)$ & 0.2201 && $b_1$ & -1.008 \\
$\sum_{k=1}^5 \varkappa^p_{k2}(0)$ & 0.4158 && $a_2$ & 0.03885 \\
$\sum_{k=1}^5 \varkappa^p_{k3}(0)$ & 0.07661 && $b_2$ & -0.1034 \\
$n_1$ & 0.7375 && $a_3$ & 0.08825 \\
$n_2$ & 3 && $b_3$ & -0.5403 \\
$n_3$ & 1 && $\Lambda$ & 0.3
\end{tabular}
\end{table}

\begin{figure}
	\center\includegraphics[width=0.85\linewidth]{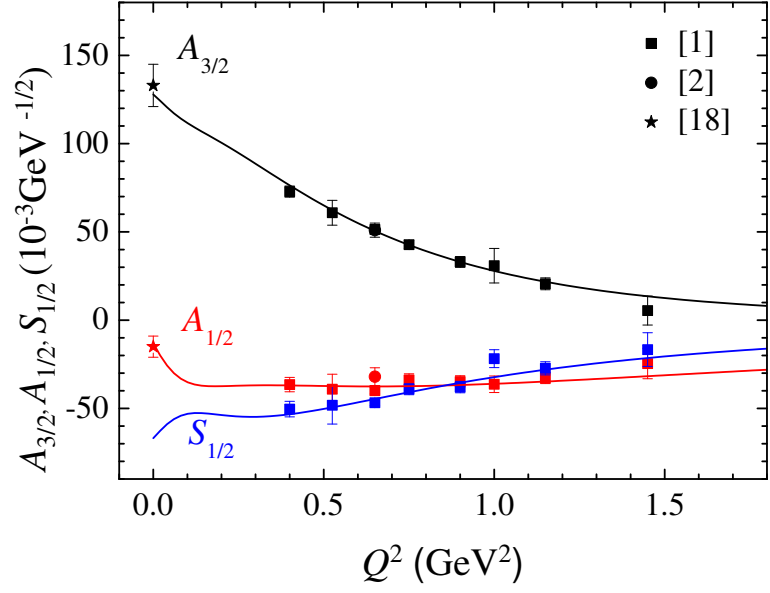}
	\caption{\label{fig:HAs}Helicity amplitudes for the transition $p\gamma^*\to N^*_+(1680)$.}
\end{figure}

\begin{figure}
	\center\includegraphics[width=0.85\linewidth]{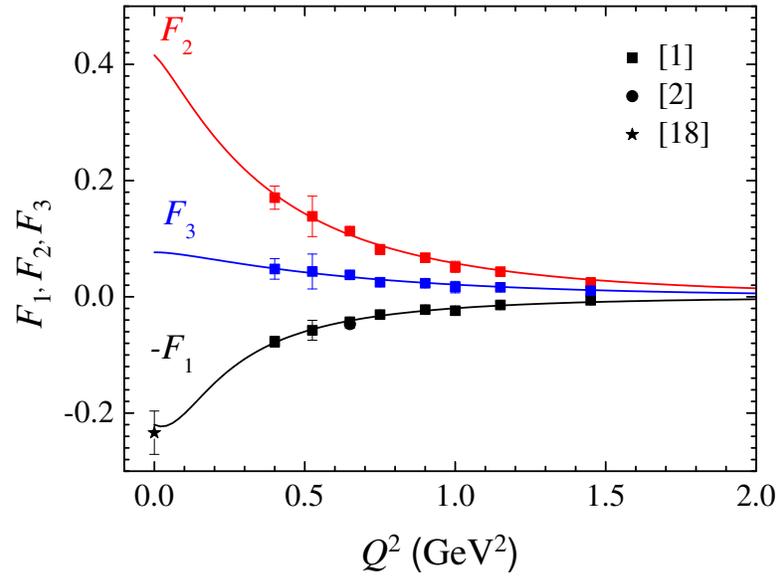}
	\caption{\label{fig:FFs}Point and gauge invariant FFs for the transition $p\gamma^*\to N^*_+(1680)$.}
\end{figure}

The dispersionlike expansions of the point and gauge invariant FFs are fitted to the experimental data on the $N^*(1680)$ electroproduction off the proton \cite{r2008PDG, 2007drechsel, az-05b} with $\chi^2/\text{DOF} = 1.05$. In the fit we use $K=5$ vector mesons---the minimal number of the mesons that is sufficient to saturate the superconvergence relations. The meson masses and the fit parameters are set out in the Tables \ref{tab:mesons} and \ref{tab:param}. The fit curves and the data points \cite{r2008PDG, 2007drechsel, az-05b} are depicted in Fig.~\ref{fig:HAs}. The figure~\ref{fig:FFs} shows the FFs $F_f(Q^2)$ extracted from the data using the formulas for the helicity amplitudes \eqref{A32}--\eqref{S12}. As can be seen in these plots, the VMD model of the point and gauge invariant FFs agrees with the data well, which supports that vector-meson dominance model is applicable to the global in $Q^2$ description of all the nucleon electromagnetic FFs and that higher-spin intrinsic symmetries are an important element in the baryon-meson field theories of the transition FFs.

It is interesting to study explicitly the ratios of the FFs, since such quantities appear to exhibit scaling behavior starting from the momentum transfers lower than $1 \text{ GeV}^2$. Such low-$Q^2$ scaling of the FF ratios has been seen in the cases of the nucleon elastic FFs \cite{2003Belitsky} and the FFs for the transition $N\to\Delta(1232)$ \cite{2009vereshkov}. However, the quality and $Q^2$-span of the available experimental data on the FFs of the $NN^*(1680)$-transition are completely insufficient to draw any definite conclusion concerning the regime of the $Q^2$-evolution of the FF ratios, although the hypothesis of the scaling behavior of the ratios does not contradict the experimental data. Further data should be anticipated to remedy this.


\section{\label{sec:conclusion}Conclusion}

In this Letter we suggested a Lagrangian of electromagnetic $NR$-inter\-actions involving spin-$\frac52$ resonances. The Lagrangian preserves all the symmetries of the free spin-$\frac52$ field and, consequently, does not modify the free-field constraints. The Lagrangian belongs to the class of consistent gauge-invariant interactions \cite{pa-ti}, which do not excite redundant lower-spin components of the RS field. Being point-invariant as it is, the Lagrangian is specific among other possible gauge-invariant Lagrangians. (1) All three terms of the minimally local Lagrangian are defined uniquely by the symmetry and their differential order. (2) The symmetry classifies the FFs in terms of the differential oder of the corresponding Lagrangian vertex. This classification is consistent with the pQCD interpretation of the FFs in the asymptotic domain.

It should be noted that these properties are also shared by the point and gauge invariant theory of the spin-$\frac32$ resonances \cite{2009vereshkov}. Besides, the point and gauge invariant $\gamma^*NR$-interactions of the spin-$\frac32$ and $\frac52$ resonances are unified in three other ways. (1) All tensor-spinor couplings are expressed through the universal tensor spinor \eqref{kernel}. The couplings of the spin-$\frac32$ are linear in the tensor spinor \eqref{kernel}, while the couplings of the spin-$\frac52$ are bilinear in it. (2) The tensor spinors \eqref{kernel} and \eqref{Gamma8} (bilinear in the coupling \eqref{kernel}) take the same place in the Lagrangians for the spin-$\frac32$ \eqref{L V} and spin-$\frac52$ \eqref{L123} resonances, respectively. (3) The pre-FF polynomials in helicity amplitudes \eqref{A32}--\eqref{S12} are the same in both cases of spin-$\frac32$ and spin-$\frac52$ resonance excitation.

All these analogies allow us to expect that this pattern remains for arbitrary high spin of the resonance. In this case the point and gauge invariance result in a unified, simple, and consistent description of all $\gamma^*NR$-interactions.


\bibliographystyle{elsarticle-num}


\end{document}